\documentclass [11pt,a4paper] {article}

\usepackage[cp1252]{inputenc}

\usepackage{amssymb}
\usepackage{amsmath}
\usepackage{amsfonts,amssymb}
\usepackage[dvips]{graphicx}
\usepackage{bbm}

\DeclareMathAlphabet{\mathpzc}{OT1}{pzc}{m}{it}

\def\SmallColSep{\setlength{\arraycolsep}{1pt}}

\newcommand*\rfrac[2]{{}^{#1}\!\!/\!_{#2}}

\begin{document}

\title{The quantum pigeonhole principle as a violation\\ of the principle of bivalence}

\author{Arkady Bolotin\footnote{$Email: arkadyv@bgu.ac.il$\vspace{5pt}} \\ \textit{Ben-Gurion University of the Negev, Beersheba (Israel)}}

\maketitle

\begin{abstract}\noindent In the paper, it is argued that the phenomenon known as the quantum pigeonhole principle (namely, three quantum particles are put in two boxes, yet no two particles are in the same box) can be explained not as a violation of Dirichlet's box principle in the case of quantum particles but as a nonvalidness of a bivalent logic for describing not-yet verified propositions relating to quantum mechanical experiments.\\

\end{abstract}

\section{Introduction}  

\noindent Do quantum systems always possess intrinsic properties? In accordance with a realist interpretation of quantum mechanics \cite{Isham}, it is appropriate to say that an individual system possesses values of its physical quantities even before these values can be measured. In this context, ``appropriate'' means that propositions asserting possession of the physical quantities can be handled using the standard propositional logic obeying \textit{the principle of bivalence} (saying that ``A proposition cannot be neither true nor false'' and ``A proposition cannot be both true and false'' \cite{Beziau}).\\

\noindent However, such an assumption brings about a violation of an abstract principle of combinatorial analysis, namely, Dirichlet's box principle also known as the pigeonhole principle \cite{Cassels}.\\

\noindent Indeed, let us consider three quantum particles and suppose that each particle has either the property $x$ or the property $\mathrm{not-}x$. Let $X_j$ denote the proposition asserting that the particle $j\in \{1,2,3\}$ possesses the property $x$ and, correspondingly, let $\neg X_j$ denote the proposition that this particle possesses the alternative property $\mathrm{not-}x$.\\

\noindent Using the language of the paper \cite{Aharonov}, one may say that the particle $j$ is in the box ``$x$'' if the particle possesses the property $x$ and analogously the particle $j$ is in the box ``$\mathrm{not-}x$'' if it has the property $\mathrm{not-}x$.\\

\noindent Let ${[\![ \diamond ]\!]}_v$, where the symbol $\diamond$ can be replaced by any proposition (compound or simple), refer to a \textit{valuation}, that is, an assignment of a truth-value $\mathfrak{v}$ to a proposition $\diamond$, explicitly,\smallskip

\begin{equation} \label{1} 
   {[\![ \diamond ]\!]}_v
   =
   \mathfrak{v} \in \{\mathfrak{v}\}
   \;\;\;\;  ,
\end{equation}
\smallskip

\noindent where $\{\mathfrak{v}\}$ is the set of the truth-values ranging from the value 0 (denoting \textit{the falsity}) to the value 1 (denoting \textit{the truth}). Furthermore, let the truth value of the negation ${[\![ \neg X_j ]\!]}_v$ be defined by the following axiom\smallskip

\begin{equation} \label{2} 
   {[\![ \neg X_j ]\!]}_v
   =
   1
   -
   {[\![ X_j ]\!]}_v
   \;\;\;\;  .
\end{equation}
\smallskip

\noindent According to the classical distributive law, for any two particles $j$ and $k$, where $j < k \le 3$, the equality must hold\smallskip

\begin{equation} \label{3} 
   \left(
   X_j  \lor \neg X_j
   \right)
   \land
   \left(
   X_k  \lor \neg X_k
   \right)
   =
   \mathrm{Same}_{jk}
   \lor
   \mathrm{Diff}_{jk}
   \;\;\;\;  .
\end{equation}
\smallskip

\noindent where\smallskip

\begin{equation} \label{4} 
   \mathrm{Same}_{jk}
   \equiv
   \left(
   X_j  \land X_k
   \right)
   \lor
   \left(
   \neg X_j  \land \neg X_k
   \right)
   \;\;\;\;  ,
\end{equation}

\begin{equation} \label{5} 
   \mathrm{Diff}_{jk}
   \equiv
   \left(
   X_j  \land \neg X_k
   \right)
   \lor
   \left(
   \neg X_j  \land X_k
   \right)
   \;\;\;\;  .
\end{equation}
\smallskip

\noindent At this point, let us consider the case of a bivalent logic with the set of the truth-values $\{\mathfrak{v}\} = \{0,1\}$.\\

\noindent In such a case, among three propositions $X_j$ there are at least two having the same truth-value, 0 or 1. This means that ahead of the verification of the propositions $X_j$, a pair of the particles is always in the same box – either ``$x$'' or ``$\mathrm{not-}x$'', which can be presented in the form of the pigeonhole principle\smallskip

\begin{equation} \label{6} 
   {[\![ 
      \,
      \mathrm{Same}_{12}
      \lor
      \mathrm{Same}_{13}
      \lor
      \mathrm{Same}_{23}
      \,
   ]\!]}_v
   =
   1
   \;\;\;\;  .
\end{equation}
\smallskip

\noindent By contrast, let us assume that the cardinality of the set of the truth-values concerning unperformed quantum mechanical experiments is not 2 but, say 3, specifically, $\{\mathfrak{v}\} = \{0,\rfrac{1}{2},1\}$ where the additional truth value $\rfrac{1}{2}$ is interpreted as ``neither true nor false'' (and 1 is the only \textit{designated truth value}).
$\,$\footnote{\label{f1}This can be Kleene's (strong) logic $K_3$ or the 3-valued {\L}ukasiewicz system \cite{Gottwald,Miller}.\vspace{5pt}}\\
 
\noindent Using the corresponding three-valued truth-table, it is straightforward to show that in the given case there are instances in which no two propositions $X_j$ and $X_k$ have the same truth-value.\\

\noindent For example, when ${[\![ X_1 ]\!]}_v=1$, ${[\![ X_2 ]\!]}_v=\rfrac{1}{2}$ and ${[\![ X_3 ]\!]}_v=0$, one can say that before the verification, the particle 1 is in the box ``$x$'', the particle 3 is in the box ``$\mathrm{not-}x$'', while the particle 2 is neither in the box ``$x$'' nor in the box ``$\mathrm{not-}x$''. In other words, in this instance, no two particles are in the same box.\\

\noindent In the said instance, as the conjunction $X_1 \land X_2$ along with the conjunction $\neg X_2 \land \neg X_3$ cannot be evaluated to the truth
$\,$\footnote{\label{f2}As long as the conjunction of two propositions is the weakest proposition among the two.\vspace{5pt}},
 ${[\![ \mathrm{Same}_{12} ]\!]}_v$ and ${[\![ \mathrm{Same}_{23} ]\!]}_v$ are not equal to 1. Thus, in case of a non-bivalent logic one must get\smallskip

\begin{equation} \label{7} 
   {[\![ 
      \,
      \mathrm{Same}_{12}
      \lor
      \mathrm{Same}_{13}
      \lor
      \mathrm{Same}_{23}
      \,
   ]\!]}_v
    \neq
   1
   \;\;\;\;  .
\end{equation}
\smallskip

\noindent As follows, a violation of the pigeonhole principle described in the paper \cite{Aharonov} (namely, three quantum particles are put in two boxes, yet no two particles are in the same box) can be viewed not as a failure of Dirichlet's box principle in the case of quantum particles but as a nonvalidness of a bivalent semantics for treating not-yet verified propositions about properties of quantum mechanical systems.\\

\noindent Let us develop this line of argument further in this paper.\\

\section{Preliminaries}  

\noindent Following the setup introduced in the paper \cite{Aharonov} let us consider the complex Hilbert space $\mathcal{H}$ of finite dimension 4, i.e., $\mathcal{H} \equiv \mathbb{C}^4$, related to the two-qubit system with each qubit (called a ``particle'') prepared in the superposition\smallskip

\begin{equation} \label{8} 
      |{\Psi}^{z+}_{j}\rangle
      =
      \frac{1}{\sqrt{2}}
      \big(
      |{\Psi}^{x+}_{j}\rangle
      +
      |{\Psi}^{x-}_{j}\rangle
      \big)
      \;\;\;\;  ,
\end{equation}
\smallskip

\noindent where $| \,\cdot\, \rangle$ are the normalized eigenvectors of the Pauli spin matrices.\\

\noindent The projection operator $\hat{P}^{z+}_{jk}$\smallskip

\begin{equation} \label{9} 
      \hat{P}^{z+}_{jk}
      \equiv
      |{\Psi}^{z+}_{j}\rangle \langle {\Psi}^{z+}_{j}|
      \otimes
      |{\Psi}^{z+}_{k}\rangle \langle {\Psi}^{z+}_{k}|
      =
      \!\left[
         \begingroup\SmallColSep
         \begin{array}{r r r r}
            1 & 0 & 0 & 0\\
            0 & 0 & 0 & 0\\
            0 & 0 & 0 & 0\\
            0 & 0 & 0 & 0
         \end{array}
         \endgroup
      \right]
      \;\;\;\;   
\end{equation}
\smallskip

\noindent corresponds to the proposition $Z^{+}_{jk}$ asserting that both particles of the bipartite composite system have the same spin angular momentum value $+\rfrac{\hbar}{2}$ along the axis $z$. Together with this, the projection operators $\hat{P}^{\mathrm{Same}}_{jk}$  and $\hat{P}^{\mathrm{Diff}}_{jk}$ corresponding to the propositions $\mathrm{Same}_{jk}$ and $\mathrm{Diff}_{jk}$ introduced in (\ref{4}) and (\ref{5}) are given explicitly by\smallskip

\begin{equation} \label{10} 
      \hat{P}^{\mathrm{Same}}_{jk}
      =
      \frac{1}{2}
      \!\left[
         \begingroup\SmallColSep
         \begin{array}{r r r r}
            1 & 0 & 0 & 1\\
            0 & 1 & 1 & 0\\
            0 & 1 & 1 & 0\\
            1 & 0 & 0 & 1
         \end{array}
         \endgroup
      \right]
      \;\;\;\;  ,
\end{equation}

\begin{equation} \label{11} 
      \hat{P}^{\mathrm{Diff}}_{jk}
      =
      \frac{1}{2}
      \!\left[
         \begingroup\SmallColSep
         \begin{array}{r r r r}
             1 &  0 &  0 & -1\\
             0 &  1 & -1 &  0\\
             0 & -1 &  1 &  0\\
            -1 &  0 &  0 &  1
         \end{array}
         \endgroup
      \right]
      \;\;\;\;  .
\end{equation}
\smallskip

\noindent As it follows from here,\smallskip

\begin{equation} \label{12} 
      \hat{P}^{\mathrm{Same}}_{jk}
      \hat{P}^{\mathrm{Diff}}_{jk}
      =
      \hat{P}^{\mathrm{Diff}}_{jk}
      \hat{P}^{\mathrm{Same}}_{jk}
      =
      \!\left[
         \begingroup\SmallColSep
         \begin{array}{r r r r}
            0 & 0 & 0 & 0\\
            0 & 0 & 0 & 0\\
            0 & 0 & 0 & 0\\
            0 & 0 & 0 & 0
         \end{array}
         \endgroup
      \right]
      \equiv
      \hat{0}
      \;\;\;\;  ,
\end{equation}

\begin{equation} \label{13} 
      \hat{P}^{\mathrm{Same}}_{jk}
      +
      \hat{P}^{\mathrm{Diff}}_{jk}
      =
      \!\left[
         \begingroup\SmallColSep
         \begin{array}{r r r r}
            1 & 0 & 0 & 0\\
            0 & 1 & 0 & 0\\
            0 & 0 & 1 & 0\\
            0 & 0 & 0 & 1
         \end{array}
         \endgroup
      \right]
      \equiv
      \hat{1}
      \;\;\;\;  ,
\end{equation}
\smallskip

\noindent where $\hat{0}$ is the zero matrix and $\hat{1}$ is the identity matrix (the operator of the identity mapping) on $\mathbb{C}^4$.\\

\noindent Let us consider a lattice $L(\mathbb{C}^4)$ of the subspaces of $\mathbb{C}^4$ in which the partial order $\le$ is set inclusion $\subseteq$, the meet $\sqcap$ is set intersection $\cap$ and the join $\sqcup$ is the internal direct sum of any pairwise disjoint sequence of the subspaces of $\mathbb{C}^4$. The lattice $L(\mathbb{C}^4)$ is bounded, with the trivial space $\{0\}$ equal to the range (column space) of the zero matrix, $\mathrm{ran}(\hat{0})=\{0\}$, as the bottom and the whole space $\mathbb{C}^4$ equal to the range of the identity matrix, $\mathrm{ran}(\hat{1})=\mathbb{C}^4$, as the top.\\

\noindent Because any subspace of $\mathbb{C}^4$ is the range of some unique projection operator on $\mathbb{C}^4$, there is a one-to-one correspondence between the subspaces of $\mathbb{C}^4$ and the corresponding projection operators. Thus, one can take the projection operators to be the elements of $L(\mathbb{C}^4)$.\\

\noindent Specifically, as\smallskip

\begin{equation} \label{14} 
      \mathrm{ran}(\hat{P}^{\mathrm{Same}}_{jk})
      \subseteq
      \mathrm{ker}(\hat{P}^{\mathrm{Diff}}_{jk})
      =
      \mathrm{ran}(\hat{1} - \hat{P}^{\mathrm{Diff}}_{jk})
      \;\;\;\;  ,
\end{equation}
\smallskip

\noindent one can define the partial order $\hat{P}^{\mathrm{Same}}_{jk} \le (\hat{1} - \hat{P}^{\mathrm{Diff}}_{jk})$ by setting $\hat{P}^{\mathrm{Same}}_{jk} \sqcap (\hat{1} - \hat{P}^{\mathrm{Diff}}_{jk}) = \hat{P}^{\mathrm{Same}}_{jk}$  which means that the meet of $\hat{P}^{\mathrm{Same}}_{jk}$  and $\hat{P}^{\mathrm{Diff}}_{jk}$  in $L(\mathbb{C}^4)$ can be defined by\smallskip

\begin{equation} \label{15} 
      \hat{P}^{\mathrm{Same}}_{jk}
      \sqcap
      \hat{P}^{\mathrm{Diff}}_{jk}
      =
      \hat{P}^{\mathrm{Same}}_{jk}
      \hat{P}^{\mathrm{Diff}}_{jk}
      =
      \hat{0}
      \;\;\;\;  .
\end{equation}
\smallskip

\noindent Since the subspaces $\mathrm{ran}(\hat{P}^{\mathrm{Same}}_{jk})$ and $\mathrm{ran}(\hat{P}^{\mathrm{Diff}}_{jk})$ are disjoint, namely\smallskip

\begin{equation} \label{16} 
      \mathrm{ran}(\hat{P}^{\mathrm{Same}}_{jk})
      \cap
      \mathrm{ran}(\hat{P}^{\mathrm{Diff}}_{jk})
      =
      \mathrm{ran}(\hat{P}^{\mathrm{Same}}_{jk}\hat{P}^{\mathrm{Diff}}_{jk})
      =
      \mathrm{ran}(\hat{0})
      =
      \{0\}
      \;\;\;\;  ,
\end{equation}
\smallskip

\noindent the join of $\hat{P}^{\mathrm{Same}}_{jk}$  and $\hat{P}^{\mathrm{Diff}}_{jk}$  in $L(\mathbb{C}^4)$ can be defined as their sum, i.e.,\smallskip

\begin{equation} \label{17} 
      \hat{P}^{\mathrm{Same}}_{jk}
      \sqcup
      \hat{P}^{\mathrm{Diff}}_{jk}
      =
      \hat{P}^{\mathrm{Same}}_{jk}
      +
      \hat{P}^{\mathrm{Diff}}_{jk}
      =
      \hat{1}
      \;\;\;\;  .
\end{equation}
\smallskip

\noindent As an immediate consequence of such definitions, it follows that $\hat{P}^{z+}_{jk} \sqcap \hat{P}^{\mathrm{Same}}_{jk}$  and $\hat{P}^{z+}_{jk} \sqcup \hat{P}^{\mathrm{Same}}_{jk}$ are not defined in $L(\mathbb{C}^4)$ because $\hat{P}^{z+}_{jk} \hat{P}^{\mathrm{Same}}_{jk} \neq \hat{P}^{\mathrm{Same}}_{jk} \hat{P}^{z+}_{jk}$ and therefore neither $\hat{P}^{z+}_{jk} \hat{P}^{\mathrm{Same}}_{jk}$  nor $\hat{P}^{\mathrm{Same}}_{jk} \hat{P}^{z+}_{jk}$ is the projection operator on $\mathbb{C}^4$ (the same concerns $\hat{P}^{z+}_{jk} \hat{P}^{\mathrm{Diff}}_{jk}$  and $\hat{P}^{\mathrm{Diff}}_{jk} \hat{P}^{z+}_{jk}$).\\

\noindent Given that the projection operator $\hat{1}$ leaves invariant any vector lying in the space $\mathbb{C}^4$, the range of $\hat{1}$, a proposition represented by $\hat{1}$ must be true in any state of the system, i.e., such a proposition must be a tautology $\top$. Also, as the projection operator $\hat{0}$ annihilates any vector in $\mathbb{C}^4$, the null space of $\hat{0}$, a proposition represented by $\hat{0}$ must be false in any state of the system, i.e., this proposition must be a contradiction $\bot$.\\

\noindent This can be written as\smallskip

\begin{equation} \label{18} 
   |{\Psi}\rangle
   \in
   \mathrm{ran}(\hat{1})
   \quad
   \implies
   \quad
   v(\hat{1})
   =
   {[\![ \top ]\!]}_v
   =
   1
   \;\;\;\;  ,
\end{equation}

\begin{equation} \label{19} 
   |{\Psi}\rangle
   \in
   \mathrm{ker}(\hat{0})
   \quad
   \implies
   \quad
   v(\hat{0})
   =
   {[\![ \bot ]\!]}_v
   =
   0
   \;\;\;\;  ,
\end{equation}
\smallskip

\noindent where the symbol $\implies$ means ``implies'' or ``if \dots then'', $v$ denotes the truth-function that maps a given projection operator to the truth value of the corresponding proposition.\\

\noindent Let $\hat{P}_A$ and $\hat{P}_B$ denote the projection operators representing the propositions $A$ and $B$. Then, to decide the truth values of disjunction, conjunction and negation of these propositions, let the following valuational axioms hold\smallskip

\begin{equation} \label{20} 
    v(\hat{P}_A \sqcup \hat{P}_B)
    =
   {[\![ A \!\lor\! B ]\!]}_v
   \;\;\;\;  ,
\end{equation}

\begin{equation} \label{21} 
    v(\hat{P}_A \sqcap \hat{P}_B)
    =
   {[\![ A \!\land\! B ]\!]}_v
   \;\;\;\;  ,
\end{equation}

\begin{equation} \label{22} 
    v(\hat{1}-\hat{P}_A)
    =
   {[\![ \neg A ]\!]}_v
   \;\;\;\;  .
\end{equation}
\smallskip

\noindent In this manner, disjunction and conjunction on the propositions $Z^{+}_{jk}$ and ${\mathrm{Same}}_{jk}$ are undefined since $\hat{P}^{z+}_{jk} \sqcap \hat{P}^{\mathrm{Same}}_{jk}$ and $\hat{P}^{z+}_{jk} \sqcup \hat{P}^{\mathrm{Same}}_{jk}$ are not defined in $L(\mathbb{C}^4)$.\\

\noindent According to such valuations, one gets\smallskip

\begin{equation} \label{23} 
    v(\hat{P}^{\mathrm{Same}}_{jk} \sqcup \hat{P}^{\mathrm{Diff}}_{jk})
    =
   {[\![
      \,
      {\mathrm{Same}}_{jk}
      \lor
      {\mathrm{Diff}}_{jk} 
      \,
   ]\!]}_v
   =
   1
   \;\;\;\;  ,
\end{equation}

\begin{equation} \label{24} 
    v(\hat{P}^{\mathrm{Same}}_{jk} \sqcap \hat{P}^{\mathrm{Diff}}_{jk})
    =
   {[\![
      \,
      {\mathrm{Same}}_{jk}
      \land
      {\mathrm{Diff}}_{jk} 
      \,
   ]\!]}_v
   =
   0
   \;\;\;\;  ,
\end{equation}
\smallskip

\noindent which means that the statement ``Either two particles are in the same box or it is not the case that two particles are in the same box'' is always true.\\

\section{The intermediate truth-value of the proposition $\mathbf{Same}_{\boldsymbol{jk}}$}  

\noindent After the preparation, the two-qubit system’s spin state is preselected in the $z+$ direction, i.e., in the state $|{\Psi}^{z+}_{jk}\rangle$\\

\begin{equation} \label{25} 
      |{\Psi}^{z+}_{jk}\rangle
      \equiv
      |{\Psi}^{z+}_{j}\rangle
      \otimes
      |{\Psi}^{z+}_{k}\rangle
      =
       \left[
         \!\!
         \begin{array}{r}
            1\\
            0
         \end{array}
         \!\!
      \right]
      \otimes
       \left[
         \!\!
         \begin{array}{r}
            1\\
            0
         \end{array}
         \!\!
      \right]
      =
      \left[
         \!\!
         \begin{array}{r}
            1\\
            0\\
            0\\
            0
         \end{array}
         \!\!
      \right]
      \;\;\;\;   
\end{equation}
\smallskip

\noindent lying in the range of the projection operator $\hat{P}^{z+}_{jk}$:\smallskip

\begin{equation} \label{26} 
   \mathrm{ran}(\hat{P}^{z+}_{jk})
   =
   \left\{
   \left.
   \left[
      \!\!
      \begin{array}{r}
         a\\
         0\\
         0\\
         0
      \end{array}
      \!\!
   \right]
   \right|
   a \in \mathbb{R}
   \right\}
   \quad
   \text{,}
   \quad
   \mathrm{ker}(\hat{P}^{z+}_{jk})
   =
   \left\{
   \left.
   \left[
      \!\!
      \begin{array}{r}
         0\\
         b\\
         c\\
         d
      \end{array}
      \!\!
   \right]
   \right|
   b,c,d \in \mathbb{R}
   \right\}
   \;\;\;\;  .
\end{equation}
\smallskip

\noindent In this state, the proposition $Z^{+}_{jk}$ is definite and has the truth-value 1. Hence, one can say that in the preselected state, the two-qubit system possesses an intrinsic property, specifically, both qubits have the same spin value $+\rfrac{\hbar}{2}$ along the $z$-axis.\\ 

\noindent Given that each projection operator leaves invariant any vector lying in its range and annihilates any vector lying in its null space, the definiteness of the proposition $Z^{+}_{jk}$ can be written down as its bivalence, i.e.,\smallskip

\begin{equation} \label{27} 
   |{\Psi}^{z+}_{jk}\rangle
   \in
   \mathrm{ran}(\hat{P}^{z+}_{jk})
   \quad
   \implies
   \quad
   v(\hat{P}^{z+}_{jk})
   =
   {[\![ Z^{+}_{jk} ]\!]}_v
   =
   1
   \;\;\;\;  ,
\end{equation}

\begin{equation} \label{28} 
   |{\Psi}^{z+}_{jk}\rangle
   \notin
   \mathrm{ker}(\hat{P}^{z+}_{jk})
   \quad
   \implies
   \quad
   v(\hat{P}^{z+}_{jk})
   =
   {[\![ Z^{+}_{jk} ]\!]}_v
   \neq
   0
   \;\;\;\;  .
\end{equation}
\smallskip

\noindent Now, consider the range and the null space of the projection operator $\hat{P}^{\mathrm{Same}}_{jk}$:\smallskip

\begin{equation} \label{29} 
   \mathrm{ran}(\hat{P}^{\mathrm{Same}}_{jk})
   =
   \left\{
   \left.
   \left[
      \!\!
      \begin{array}{r}
         a\\
         b\\
         b\\
         a
      \end{array}
      \!\!
   \right]
   \right|
   a,b \in \mathbb{R}
   \right\}
   \quad
   \text{,}
   \quad
   \mathrm{ker}(\hat{P}^{\mathrm{Same}}_{jk})
   =
   \left\{
   \left.
   \left[
      \!\!
      \begin{array}{r}
         -d\\
         -c\\
          c\\
          d
      \end{array}
      \!\!
   \right]
   \right|
   c,d \in \mathbb{R}
   \right\}
   \;\;\;\;  .
\end{equation}
\smallskip

\noindent Comparing (\ref{25}) with (\ref{29}) makes it evident that the preselected vector $|{\Psi}^{z+}_{jk}\rangle$ does not lie in the range of the projection operator $\hat{P}^{\mathrm{Same}}_{jk}$ or in its null space. For that reason, one can assert that in the state $|{\Psi}^{z+}_{jk}\rangle$ the truth-value of the proposition $\mathrm{Same}_{jk}$ cannot be 1 or 0, that is, $\mathrm{Same}_{jk}$ does not obey the principle of bivalence, explicitly,\smallskip

\begin{equation} \label{30} 
   |{\Psi}^{z+}_{jk}\rangle
   \notin
   \mathrm{ran}(\hat{P}^{\mathrm{Same}}_{jk})
   \quad
   \implies
   \quad
   v(\hat{P}^{\mathrm{Same}}_{jk})
   =
   {[\![ {\mathrm{Same}}_{jk} ]\!]}_v
   \neq
   1
   \;\;\;\;  ,
\end{equation}

\begin{equation} \label{31} 
   |{\Psi}^{z+}_{jk}\rangle
   \notin
   \mathrm{ker}(\hat{P}^{\mathrm{Same}}_{jk})
   \quad
   \implies
   \quad
   v(\hat{P}^{\mathrm{Same}}_{jk})
   =
   {[\![ {\mathrm{Same}}_{jk} ]\!]}_v
   \neq
   0
   \;\;\;\;  .
\end{equation}
\smallskip

\noindent Expressed differently, in the intermediate state that exists after the preparation but before the (strong and simultaneous) measurement of particles' spins along the $x$-axis (that is to say, particles' presence in the boxes ``spin $x+$'' and ``spin $x-$'') the statement ``Two particles are in the same box'' is neither true nor false.\\

\noindent Next, consider the disjunction ${\mathrm{Same}}_{12} \lor {\mathrm{Same}}_{13}$: As stated by the valuational axiom (\ref{20}), its intermediate truth-value is determined by the join of the projection operators $\hat{P}^{\mathrm{Same}}_{12}$ and $\hat{P}^{\mathrm{Same}}_{13}$ in $L(\mathcal{C}^4)$\smallskip

\begin{equation} \label{32} 
   v(
      \hat{P}^{\mathrm{Same}}_{12}
      \sqcup
      \hat{P}^{\mathrm{Same}}_{13}
   )
   =
   {[\![
       \,
       {\mathrm{Same}}_{12}
       \lor
       {\mathrm{Same}}_{13} 
       \,
   ]\!]}_v
   \;\;\;\;  ,
\end{equation}
\smallskip

\noindent which is given by $\hat{P}^{\mathrm{Same}}_{12} \sqcup \hat{P}^{\mathrm{Same}}_{13} = \hat{P}^{\mathrm{Same}}_{13}$ consistent with the set  inclusion $\mathrm{ran}(\hat{P}^{\mathrm{Same}}_{12}) \subseteq \mathrm{ran}(\hat{P}^{\mathrm{Same}}_{13})$.\\

\noindent Thus, it must be\smallskip

\begin{equation} \label{33} 
   v(\hat{P}^{\mathrm{Same}}_{jk})
   =
   {[\![ 
      \,
      \mathrm{Same}_{12}
      \lor
      \mathrm{Same}_{13}
      \lor
      \mathrm{Same}_{23}
      \,
   ]\!]}_v
   \;\;\;\;   
\end{equation}
\smallskip

\noindent since\smallskip

\begin{equation} \label{34} 
   \left(
      \hat{P}^{\mathrm{Same}}_{12}
      \sqcup
      \hat{P}^{\mathrm{Same}}_{13}
   \right)
   \sqcup
   \hat{P}^{\mathrm{Same}}_{23}
   =
   \hat{P}^{\mathrm{Same}}_{13}
   \sqcup
   \hat{P}^{\mathrm{Same}}_{23}
   =
   \hat{P}^{\mathrm{Same}}_{23}
   \;\;\;\;   
\end{equation}
\smallskip

\noindent in agreement with\smallskip

\begin{equation} \label{35} 
   \mathrm{ran}
   \left(
      \hat{P}^{\mathrm{Same}}_{12}
      \sqcup
      \hat{P}^{\mathrm{Same}}_{13}
   \right) 
   \subseteq
   \mathrm{ran}
   \left(
      \hat{P}^{\mathrm{Same}}_{23}
   \right)
   \;\;\;\;    .
\end{equation}
\smallskip

\noindent From (\ref{33}) it immediately follows that the pigeonhole principle does not hold in the intermediate state of the quantum particles, that is,\smallskip

\begin{equation} \label{36} 
   {[\![ 
      \,
      \mathrm{Same}_{12}
      \lor
      \mathrm{Same}_{13}
      \lor
      \mathrm{Same}_{23}
      \,
   ]\!]}_v
   \neq
   \{ 0,1 \}
   \;\;\;\;  .
\end{equation}
\smallskip

\section{Concluding remarks}  

\noindent In logical terms, the pigeonhole principle boils down to the statement that among three propositions $X_1$, $X_2$ and $X_3$ there exist at least two that have the same bivalent truth-value, i.e., $0$ or $1$. Consequently, the disjunction of the set of three logical connectives $\mathrm{Same}_{jk} \stackrel{\text{\tiny def}}{=} X_j \!\!\!\iff\!\!\! X_k$ (where the symbol $\iff$ denotes ``equivalent'' and $j \neq k$) must always have the value $1$.\\

\noindent In the paper \cite{Aharonov} it is suggested that a quantum violation of the pigeonhole principle is an indication that Dirichlet's box principle (which ``encapsulates abstract mathematical notions that go to the core of what numbers and counting are, so it underlies, implicitly or explicitly, virtually the whole of mathematics'') does not hold in the case of quantum particles.\\

\noindent But, as it has been just demonstrated in the presented paper, the quantum violation of the pigeonhole principle may have another, ``less dramatic'', so to speak, explanation: It can be a sign that a logic defined as the relations between projection operators associated with quantum particles does not obey the principle of bivalence.\\

\bibliographystyle{References}
\bibliography{References_Minor_Revision}

\end{document}